# Illinois Accelerator Research Center

Thomas K. Kroc[a], Charlie A Cooper[a,1]

[a]*Illinois Accelerator Research Center, Fermi National Accelerator Research Lab, PO Box 500, Batavia, Illinois 60510, USA*

**Abstract**

The Illinois Accelerator Research Center (IARC) hosts a new accelerator development program at Fermi National Accelerator Laboratory. IARC provides access to Fermi's state-of-the-art facilities and technologies for research, development and industrialization of particle accelerator technology. In addition to facilitating access to available existing Fermi infrastructure, the IARC Campus has a dedicated 36,000 ft2 heavy assembly building (HAB) with all the infrastructure needed to develop, commission and operate new accelerators. Connected to the HAB is a 47,000 ft2 Office, Technology and Engineering (OTE) building, paid for by the state, that has office, meeting, and light technical space. The OTE building, which contains the Accelerator Physics Center, and nearby Accelerator and Technical divisions provide IARC collaborators with unique access to world class expertise in a wide array of accelerator technologies. At IARC scientists and engineers from Fermilab and academia work side by side with industrial partners to develop breakthroughs in accelerator science and translate them into applications for the nation's health, wealth and security.

## 1. IARC – Concept

The creation of the Illinois Accelerator Research Center (IARC) was made possible through the combined efforts of the Fermi National Accelerator Laboratory (Fermilab), the Department of Energy – Office of Science, and the State of Illinois. Together they have created a space and an organization that can help move state-of-the-art accelerator technology from the science lab into the marketplace with greater efficiency and maximize the return on investment of tax dollars.

### 1.1. Fermilab

Fermilab is, as its full name suggests, the nation's accelerator laboratory for high energy physics. It's staff of 350 accelerator scientists and engineers plus 300 technical staff design, build, and operate high-energy, high-power accelerators that must operate reliably to support an ambitious discovery-physics program. With its proximity to Argonne National Laboratory (less than 30 km), another national laboratory with extensive accelerator expertise, the area has the highest concentration of accelerator experts in the world.

### 1.2. Stewardship

The creation of IARC was first proposed to Illinois' Department of Commerce and Economic Opportunity (DCEO) in 2007. Funding was received in 2011 and the DOE donated the Heavy Assembly Building (HAB) and provided for its refurbishment. During this interval, the Accelerators for America's Future workshop was held and the report produced. Soon after, a concern was expressed by the Senate Energy and Water subcommittee in 2011 (Feinstein, 2011). The committee noted the length of time that it traditionally takes to translate breakthroughs in accelerator science and technology into applications with more direct benefits to society. One of the goals of IARC is to ease and quicken those translations. This requires moving beyond "Build accelerators and they will come," to looking for new applications for emerging accelerator technology and engaging and educating potential users.

---

[1] Corresponding author. Tel.: +1-630-840-6955; fax: +1-630-840-5307.
*E-mail address:* kroc@fnal.gov

## 2. IARC – Physical Description

The physical presence of IARC is a combination of new construction, for Office, Technology, and Engineering (OTE) and the repurposing of the heavy assembly building (HAB) that formerly housed the construction and maintenance the 5000 ton CDF detector for Fermilab's Tevatron program. The HAB provides 36,000 ft$^2$ of industrial space to develop, commission and operate new accelerators. The OTE provides 47,000 ft$^2$ of office, meeting and light technical space. Some of the space in both buildings is available to strategic partners in instances where the unique facilities available at IARC are not available elsewhere. As seen in Figure 1, IARC is located very close to Fermilab's Technical Division which provides easy access to the lab's superconducting accelerator and magnet expertise.

The HAB infrastructure includes 50 ton crane coverage, 2 MW of installed power, 1.5 MW of installed cooling capacity and a standalone helium satellite refrigeration system with a capacity of 600 W or 125 l/hr at 4K. Presently about half of the HAB including the cryogenic system is being used to assemble and test solenoids for a new experiment at Fermilab. Once that project is complete, the cryogenic system will be available for IARC project. In addition to providing space to allow others to bring new ideas to IARC for development, a couple areas are already being developed to enable the development of new applications of existing technologies. Namely, these are electron sources and electron and photon beams.

*2.1. HBESL*

The High Brightness Electron Source Lab (HBESL) is being constructed in partnership with Northern Illinois University (NIU). It will allow development and testing of new thermionic, field emission, and photo-stimulated cathodes for electron sources. Separate laser and electron source areas are being constructed for NIU to furnish.

*2.2. $A^2D^2$*

The Accelerator Application Development and Demonstration Center will employ a 2.25 kW electron source at energies up to 10 MeV to allow investigation into new applications of electron beams. Lab scale testing of new ideas can be performed. The source is a repurposed medical linac that will be available in electron or photon mode. Samples of order 1 kg or less, including liquids or gases, will be able to be irradiated. Preliminary studies of materials properties in x-ray fields can be studied.

## 3. New Technologies

A major emphasis at IARC is the development of a compact superconducting RF (SRF) accelerator. This accelerator will bring to the market many advantages over existing systems: continuous wave (CW) operation, greater operating efficiency, compact size, and high power at energies up to 10 MeV. Existing industrial accelerators are incremental adaptions of normal conducting accelerator designs for basic science from over 50 years ago. Superconducting RF acceleration is now the standard technology in use for basic science research today and is the first technology in 25 years that is capable of CW operation. CW operation of accelerators is desirable in most industrial applications. The duty factor of most normal conducting machines equates to wasted time and potential. In cases where the machine is a photon source for systems using particle of photon detectors, the time scale of CW operations is much more detector friendly. The use of new materials and heat transfer methods allows a cryogen-free system which eliminates much of the complexity of a cryogenic system and allows a much more compact design. In CW operation, peak power and average power are the same.

*3.1. Surperconducting RF (SRF)*

Five principle developments have been brought together to bring about this design of the SRF accelerator:

- $Nb_3Sn$ Coated SRF Cavities – recently developed techniques for coating niobium with tin enable operation with very high efficiency and very low heat generation.
- Low Heat Leak RF Power Couplers – New low heat leak designs eliminate the need for copper plating and minimize another source of heat that would need to be removed.
- Integrated Electron Gun – This also limits the heat that leaks into the system and allows for a compact design by eliminating the traditional transfer line between the gun and accelerator.
- Conduction Cooling and Cryocoolers – The low heat generation allows the use of cryocoolers. These along with the efficiency of conduction cooling eliminates the need for liquid cryogens.
- Low Cost RF Power Source – recent developments by Fermilab engineers enables the use of low cost magnetrons to provide RF power at costs up to 5 times less than present sources.

The combination of these technologies results in a CW 10 MeV accelerator of about 250 kW. Its size - less than a 20' cargo container, and cost - less than US$1M, make it affordable and scalable by adding more units. The SRF accelerator is envisioned to have applications in many areas including energy, the environment, security, and infrastructure.

*3.2. SRF Applications*

The development of the compact SRF accelerator opens many new application areas. We envision a new application sector for electron beams. Chemical reactions can be stimulated by promoting individual molecules above a reaction's interaction threshold. The ability to facilitate these reactions in flow processes with only a small amount of product reacting at any point in time may be advantageous compared to processes that rely on bulk heating for the reaction. Some applications are scale-ups of existing applications, some will be the fulfillment of proposed applications that have not succeeded till now, and others will be entirely new.

*3.2.1. Security*

CW operation is very desirable in this area as most interrogation techniques for contraband and special materials are detector based. Detectors are optimized for continuous but moderate flux photon fields. Energy and current can be modulated to fulfill the needs of new detector developments. Its compact size and low power requirements allow mobile installations.

*3.2.2. Environment*

Using the electron beam of photon beam to kill pathogens, break up complex organic molecules, kill pests now becomes economical with the high efficiency operation of the linac. Industrial wastewater can either be treated on-site at large manufacturing facilities or at remote sites as the accelerator is compact enough to be mobile, even taking into account the need for a power source.

*3.2.3. Energy*

The use of efficient e-beams may have applications in energy production such as petroleum upgrading, liquefying gaseous products at the well head improving recovery percentages. This would allow the capture of natural gas that is presently being flared off, wasting billions of dollars a year. The size of steam reformation plants may be dramatically reduced both in capital and operational costs. Flue gas treatment to remove sulfur and nitrogen oxides has been pursued since the 1980's. Limitations in beam power and machine reliability have hindered its development. Accelerator technology has now matured to the point where it is able to tackle such applications.

*3.2.4. In-situ environmental remediation*

The compact SRF accelerator is small enough for three or four to fit into a semi-trailer. This enables bringing the power of ionizing radiation to sites needing environmental remediation rather than having to bring materials to a treatment site.

*3.2.5. Pavements*

Each year, the US grinds away billions of dollars-worth of asphalt roadway and reapplies it. A cross-linkable binder, instead of bitumen, or as an additive to bitumen, would allow in-situ treatment once new pavement had been laid. A mobile accelerator would be driven over the pavement after each layer was laid and cross-link the binder into a more durable, resilient surface. Even a one-year life extension could represent as much as a 20% improvement in pavement lifetime and a corresponding reduction in life-time cost. Similar results could be found in pavement repair by dramatically improving the bond between the old and new materials, even in cold weather.

## 4. Other Activities

While the ultimate goal is to bring accelerator technology directly to industry, IARC needs to demonstrate that it is a viable partner. This requires showing that IARC can work within industry cost and schedule frameworks, that agreements can be set up quickly and that IARC can deliver its promised results. Building such a reputation takes time. In the interim, we are finding that other government agencies can provide a useful interface. Many technologies described above are beyond the basic R&D phase but not yet ready to be incorporated into marketable items. Other agencies have interests that are not necessarily based on financial return-on-investment. Being able to patch a damaged runway at a forward operating base with a durable patch in a short amount of time meets a critical need where the economics are of secondary consideration. This provides a feasibility demonstration that industry can see and can then work with to develop into a marketable product. To this end, IARC is responding to a number of government calls for developing new technologies and applications of accelerators.

To date IARC has participated and succeeded in obtaining an DOE Accelerator Stewardship awards for Energy and Environment. Here IARC and partners are working with the Water Reclamation District of Greater Chicago to investigate the treatment of sludge. IARC has also received a National Nuclear Security Administration award, the EERE Lab Small Business Voucher program, and had a team participate in DOE's LabCorps program with a proposal on additive manufacturing.

Superconducting magnet technology developed at Fermilab also provides opportunities for applications to the issues of today. It is working with the Clean Energy Trust to organize workshops with industry for wind power. This strives to incorporate superconducting magnet designs into efficient generators for 10 MW class turbines.

## 5. Conclusion

In the few years of its existence, IARC has made great strides in engaging industry to quickly move today's cutting edge accelerator technology from the nation's science labs to the marketplace. It continues to look for new partnerships to expand its impact.

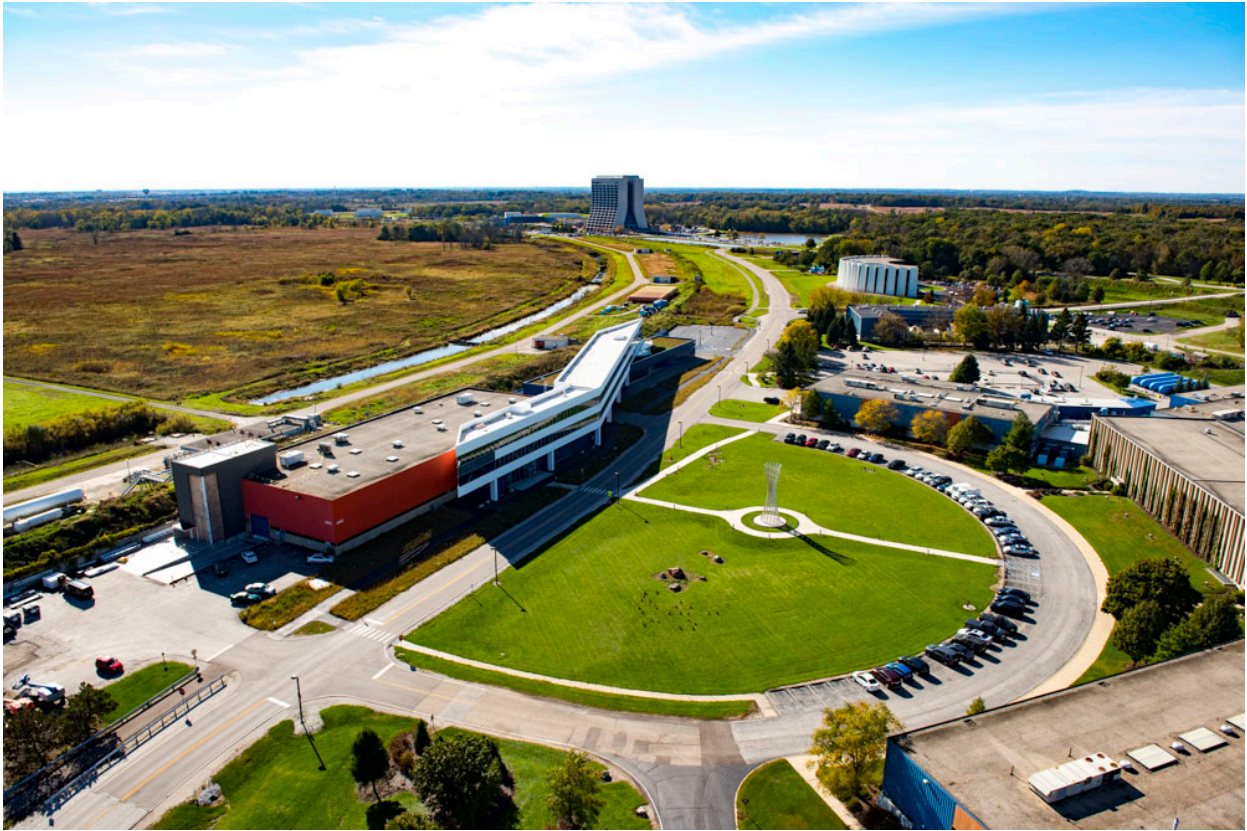

Figure 1- The IARC complex comprised of the HAB (orange) and OTE (white) buildings are near Fermilab's Technical Division (lower right). The lab's central office building (Wilson Hall) is visible on the horizon.

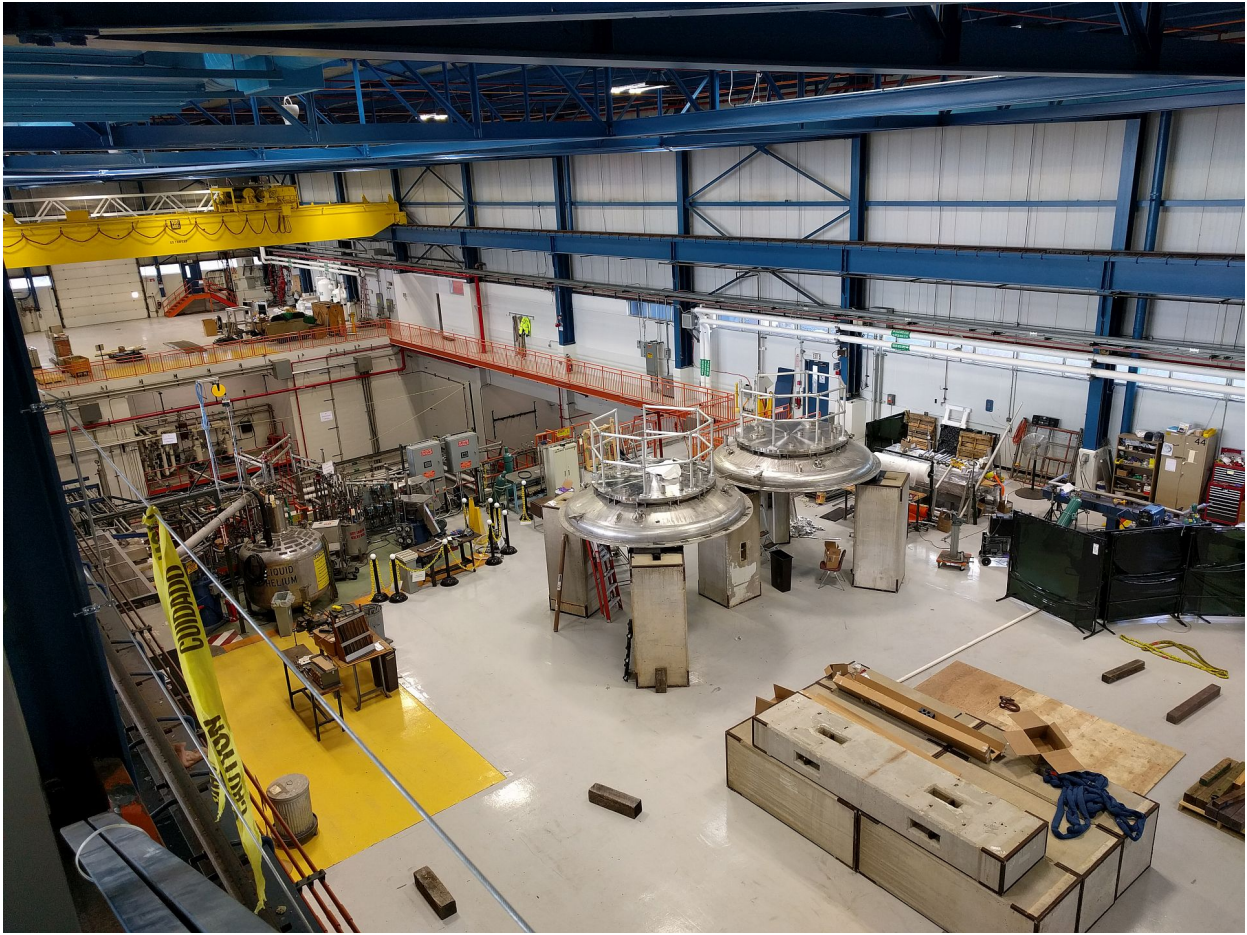

Figure 2 - Interior of the HAB. The cryogenic plant can be seen in the center left. Approximately 6,000 ft2 on either side of the center pit

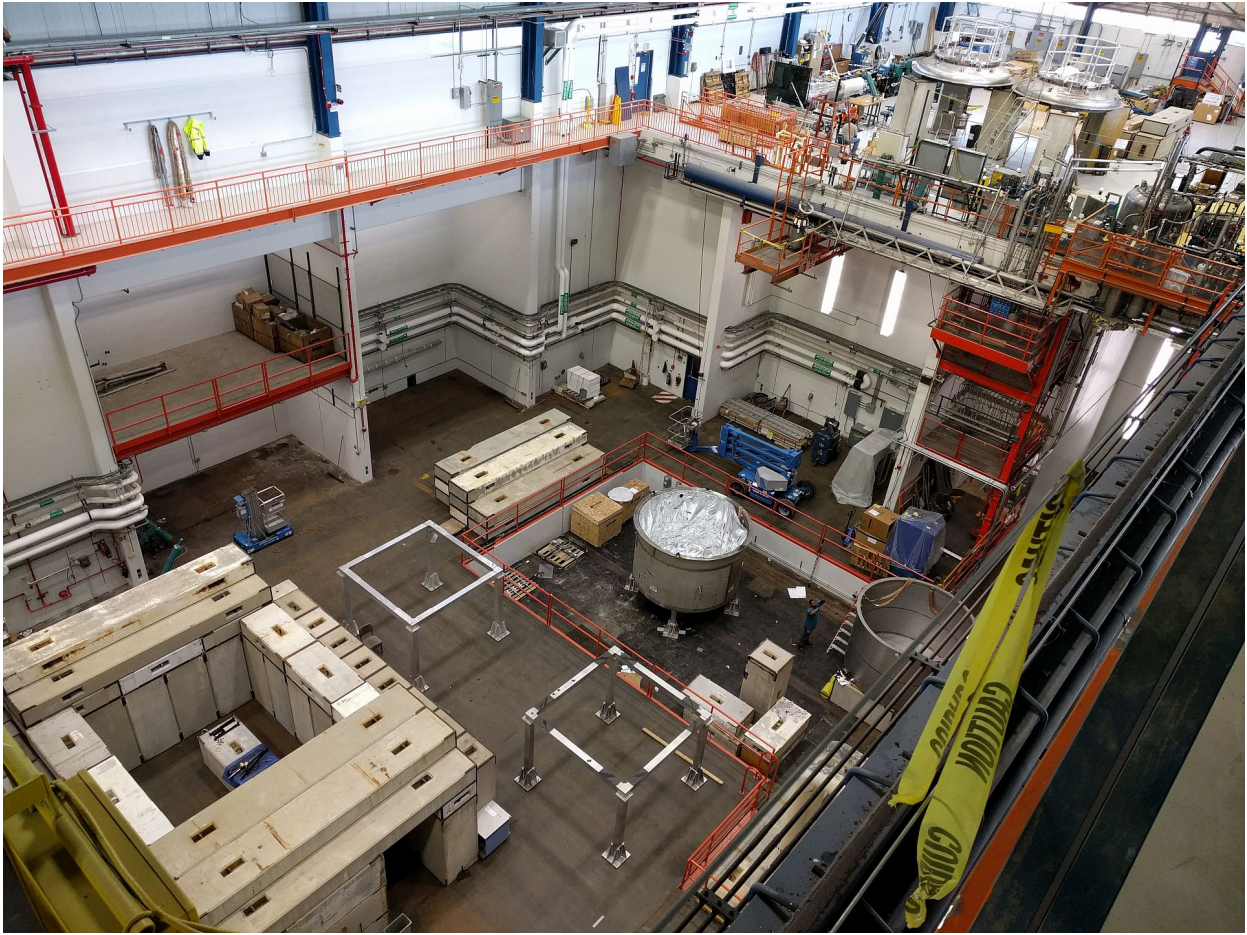

Figure 3 - The center section of the HAB with approximately 7,500 ft2 of floor space. The HBESL will occupy the two alcoves in the upper left. The left alcove will house the laser system for photo-stimulated cathodes. The right alcove will house the gun. The shield block enclosure in the lower-left will house the $A^2D^2$ electron accelerator.

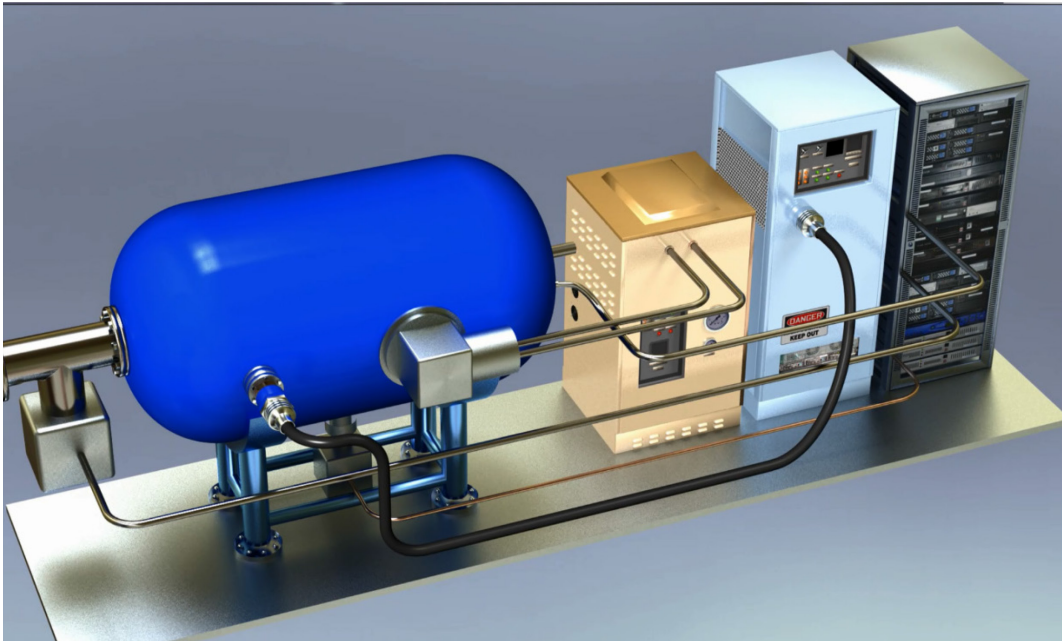

Figure 4 - The Compact SRF accelerator with physical dimensions of approximately 2x2x5 m3.